\documentclass[aps,prb,reprint,floatfix]{revtex4-2}
\usepackage{graphicx}
\usepackage{color}
\usepackage{amsmath}
\usepackage{amsthm}
\usepackage{amssymb}
\usepackage{physics}
\usepackage{cases}
\usepackage{mathrsfs}
\usepackage{subfigure}
\usepackage{comment}
\usepackage{hyperref}
\hypersetup{
	citecolor = blue,
	colorlinks = true,
	urlcolor = blue
}
\usepackage{physics2}
\usephysicsmodule{ab}

\newcommand{\al}{\alpha}

\newcommand{\si}{\sigma}

\newcommand{\p}{\prime}
\newcommand{\del}{\partial}

\newcommand{\lb}{\left(}
\newcommand{\rb}{\right)}

\newcommand{\LB}{\left[}
\newcommand{\RB}{\right]}
\newcommand{\Lb}{\langle}
\newcommand{\Rb}{\rangle}

\newcommand{\up}{\uparrow}
\newcommand{\down}{\downarrow}

\newcommand{\vv}[1]{\boldsymbol{#1}}

\begin{document}
\title{Magnon-mediated superconductivity on ferromagnetic wallpaper fermions}

\author{Koki Mizuno}
\affiliation{Department of Physics, Nagoya University, Nagoya 464-8602, Japan}
\author{Ai Yamakage}
\affiliation{Department of Physics, Nagoya University, Nagoya 464-8602, Japan}

\date{\today}

\begin{abstract}
    We study two-dimensional superconductivity mediated by magnetic fluctuations at the interface between a ferromagnetic insulator and the nonsymmorphic topological crystalline insulator with a fourfold-degenerate Dirac point, wallpaper fermion.
    We demonstrate that BCS pairing with zero center-of-mass momentum induces chiral $p$-wave superconductivity, and the Amperean pairing with center-of-mass momentum $2k_{\rm F}$ can give rise to a parity-mixed superconducting state.
    We find that the Amperean pairing exhibits a mixture of $s$-wave and $p$-wave components due to the multiband nature of the wallpaper fermion and the easy-axis anisotropy of the ferromagnetic insulator.
    Additionally, we find that the stability of the superconducting state with BCS and the Amperean pairing is governed by the easy-axis anisotropy.
\end{abstract}
\maketitle

\section{Introduction}
Heterostructures of a magnetic insulator and a topological insulator (TI) provide a platform for various intriguing phenomena, inducing the half-quantized Hall effect and spin pumping \cite{half_int_Hall, xu2014observation_half_hall, checkelsky2014trajectory_HH_FM, baker2015spinpump, jamali2015giant_spinpump},
where a single massless Dirac fermion emerging on the surface of TI plays a key role \cite{Kane_Mele_TO, Kane_Mele_graphene, Fu_Kane_3DTI, Moore_Balents, Roy2009-os, Hansen_Kane, Qi_Liang_Zhang, Tanaka_Sato_Nagaosa, Ando_topo}.
In contrast, topological crystalline insulators (TCIs) \cite{Shiozaki_crystaline_TI_TIS, shiozaki2017topological, Wang2016-uc, Liu_nonsymmorphic_TCI, Bernevig_KHgSb_nonsymmorphic, Kruthoff_band} that are protected by two glide symmetries and time-reversal symmetry (TRS), can host surface states characterized by double massless Dirac fermions with a fourfold-degenerate Dirac point, referred to as wallpaper fermions \cite{wieder2018wallpaper, glide_high_order_TI_2021, hwang2023magnetic, Kondo_insulator_PuB4, wieder2018wallpaper, Hall_WPF}.
As a result, magnetic heterostructures involving wallpaper fermions exhibit qualitatively distinct properties compared to those involving a single Dirac fermion on a TI.
For instance, such heterostructures with (anti)ferromagnetic insulators [(A)FMI] can give rise to novel (spin) Hall effects, as we have previously reported \cite{Hall_WPF}.
On the other hand, in the present paper, we focus on the emergence of superconductivity in these heterostructures.

Interfacial superconductivity mediated by the magnon-electron interaction, referred to as magnon-mediated superconductivity (MMSC), has been studied for various systems, including (anti)ferromagnetic, ferrimagnetic, skyrmionic insulators, and twisted bilayer graphene \cite{MagnonSC_Kargarian, MagnonSC_Erlandsen, Maeland_2023_Skyr, Lifetime_magnonSC_Maeland, sun2023stability, TBG_MMSC, Odd_freq_MMSC}.
In heterostructures composed of a normal metal and a ferromagnetic insulator, Cooper pairs typically form with zero center-of-mass momentum, in accordance with the conventional BCS theory \cite{FM_NM_FM_MMSC, wu2011magnon}.
However, in heterostructures involving a TI and a ferrimagnetic insulator, the magnon-mediated current-current interaction can lead to the formation of Cooper pairs with finite center-of-mass momentum, a phenomenon known as Amperean pairing \cite{MagnonSC_Kargarian}.
Moreover, the gap function associated with Amperean pairing exhibits the $p$-wave symmetric structure around the finite net momentum.
Consequently, the MMSC can enable diverse control over the superconducting state by selecting appropriate electronic systems and magnetic insulators.

In this study, we clarify the symmetry of gap functions with BCS-type pairing and Amperean pairing in the heterostructure interface between an FMI and a nonsymmorphic TCI with wallpaper fermions.
We identify a finite-temperature chiral $p$-wave superconductivity solution associated with BCS-type pairing in an isotropic-band limit.
On the other hand, we find a solution to the Amperean pairing with parity mixing.
The mixing of $p$- and $s$-wave components is enhanced by the anisotropy of wallpaper fermions.
Moreover, the easy-axis anisotropy of the FMI further amplifies this parity mixing.
These findings contrast that only the Amperean pairing with $p$-wave symmetry is allowed for MMSC at the interface of the TI and the FMI.

This paper is organized as follows.
In Sec.~\ref{sec_Model}, we derive the effective two-body interaction between wallpaper fermions mediated by ferromagnetic magnons.
Next, we derive the linearized Eliashberg equation for the BCS-type pairing with Cooper pairs without net momentum in Sec.~\ref{sec_BCS}.
In this section, we clarify that the solution of the gap equation must have a chiral $p$-wave structure with an isotropic band limit.
Next, we consider the Amperean pairing and solve the Eliashberg equation in Sec.~\ref{sec_Amp}.
Section~\ref{sec_Amp_BCS} shows the phase diagram with respect to the chemical potential and magnetic anisotropy of FMI of the superconducting state, which is the BCS and Amperan pairing state.
Section \ref{sec_Cond} summarizes the paper.

\section{Model}\label{sec_Model}
We first show model Hamiltonians for the wallpaper fermions, magnons, and their interactions. 

\subsection{Wallpaper fermion}
\label{sec:wf}
We start with a three-dimensional TCI with wallpaper fermions proximity coupled to a two-dimensional FMI.
The interface lies within the $xy$ plane, and the easy axis of the FMI is parallel to the $z$ axis.
A tight-binding Hamiltonian in the semi-infinite space, $z = a_z l\leq 0$, $l = 0, -1, \cdots$ with $a_z$ the lattice constant, is given by
\begin{align}
 H
 = \sum_{\vv{k}\in \rm{BZ}}\sum_{ll'} c_{\boldsymbol{k},l}^\dag 
 t_{ll'}(\boldsymbol{k}) c_{\boldsymbol{k}, l^\p},
\end{align}
where $\boldsymbol{k}=(k_x, k_y)$ is defined within the two-dimensional projected Brillouin zone (BZ). 
The Schr\"odinger equation on the point $\bar M$ point, $\boldsymbol{\bar M} = (\pi/a,\pi/a)$, on which the wallpaper fermions are fourfold degenerate, is given by
\begin{align}
 H(\boldsymbol{\bar M}) 
 \ket{\lambda}
 = E_\lambda 
 \ket{\lambda}.
\end{align}
This equation reduces to the eigenvalue problem
\begin{align}
 \sum_{l^\p}t_{ll^\p}(\boldsymbol{\bar M}) \eta_{l^\p\lambda} = E_\lambda \eta_{l\lambda},
\end{align}
for $\ket{\lambda} = \sum_l \eta_{l\lambda} c_{\boldsymbol{\bar M},l}^\dag \ket{0}$. 
The solution is classified into the surface states $\lambda = \lambda_{\rm s}$, which we are interested in, and the bulk states $\lambda = \lambda_{\rm b}$.
Using the basis $\eta_{l\lambda}$, $\tilde{c}_{\boldsymbol{k}\lambda} = \sum_{l}\eta_{l\lambda}^\dag c_{l\boldsymbol{k}}$, we obtain
\begin{align}
 H
 &\approx 
 \sum_{\vv{k}\in \Omega}\sum_{ll'}
 c_{\boldsymbol{k}, l}^\dag
 \ab[
 t_{ll'}(\boldsymbol{\bar M}) + \boldsymbol{v}_{ll'} \cdot \boldsymbol{k}
 ] 
 c_{\boldsymbol{k}, l'}
 \notag\\&
 \approx
 \sum_{\vv{k}\in \Omega}\sum_{\lambda_{\rm s}}
 E_{\lambda_{\rm s}}
 \tilde{c}_{\boldsymbol{k}\lambda_{\rm s}}^\dag 
 \tilde{c}_{\boldsymbol{k}\lambda_{\rm s}}
 + 
 \sum_{\vv{k}\in \Omega}\sum_{\lambda_{\rm s}\lambda_{\rm s}'}
 \tilde{c}_{\boldsymbol{k}\lambda_{\rm s}}^\dag 
 H_{\rm surf}^{\lambda_{\rm s}\lambda_{\rm s}'}(\boldsymbol{k})
 \tilde{c}_{\boldsymbol{k}\lambda_{\rm s}'},
 \label{eq_surf_Ham}
\end{align}
with the velocity matrix $\boldsymbol{v}_{ll'} = \partial t_{ll'}(\boldsymbol{k})/\partial \boldsymbol{k} |_{\boldsymbol{k}=\boldsymbol{k}_s}$.
$\boldsymbol{k}$ is measured from the $\bar M$ point. 
The sum is restricted in the region $\Omega$ near the $\bar M$ point. 
Here, $4 \times 4$ matrix $H_{\rm surf}(\boldsymbol{k})$ is an effective Hamiltonian for the surface states.
Hereafter we redefine the energy origin such that $E_{\lambda_s}=0$.

\begin{figure}
  \centering
  \includegraphics[scale=0.3]{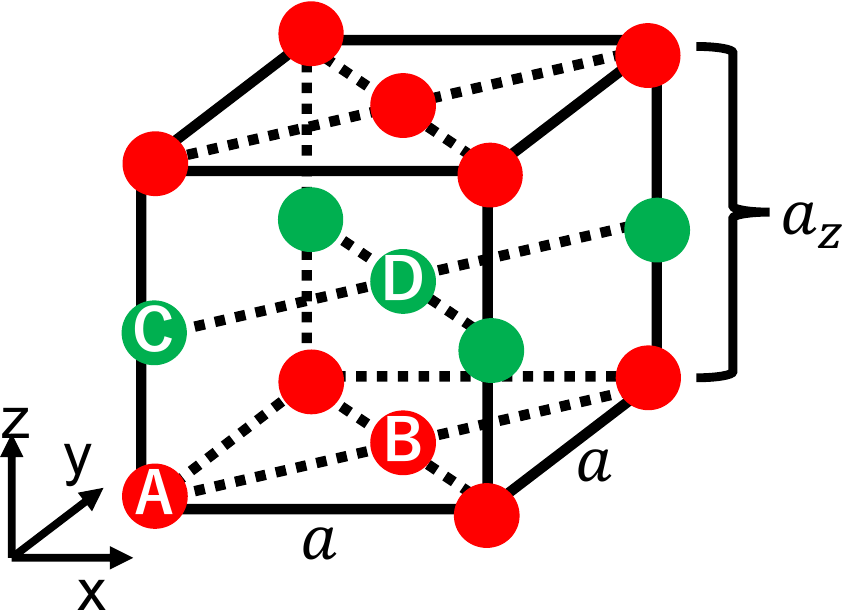}
  \caption{Crystal structure of the lattice model. 
  	This crystal has the symmetry of space group $P4/mbm$ (No.~127).
  }
  \label{SG_55}
\end{figure}

The explicit form for the wallpaper fermions is written as \cite{Hall_WPF}
\begin{align}
H_{\rm{wp}} &= \sum_{\vv{k}} c^{\dag}_{\vv{k}}H_{\mathrm{wp}}(\vv{k})c_{\vv{k}},
        \\
	H_{\mathrm{wp}}(\vv{k})
	&= 
  {-}\frac{v_{s1}}{2}\tau_x(k_x\si_x - k_y\si_y) 
    +\frac{v_{s2}}{2}\tau_z(k_x\si_x + k_y\si_y) 
    \notag\\
    &\quad
    +\frac{v_{s2}^{\p}}{2}\tau_0(k_x\si_y - k_y\si_x) -\mu\tau_{0}\sigma_{0},
    \label{WP_eff}
\end{align}
which corresponds to $H_{\rm surf}^{\lambda_{s}\lambda_{s}'}(\vv{k})$ in Eq.~(\ref{eq_surf_Ham}).
In the above equation, $\vv{k}$ is the wavevector measured from the $\bar{M}$ point in the surface BZ and $\sigma_\nu$ and $\tau_\nu$ $(\nu=x,y,z)$ are Pauli matrices corresponding to the spin and inplane sublattice [A and B (C and D)] on the surface (Fig.~\ref{SG_55}).
The identity matrices are denoted by $\sigma_0$ and $\tau_0$. 
The operator $c_{\vv{k}}$ is the four-component annihilation operator of electrons with the spin and sublattice degrees of freedom.
Hereafter, we set $a=1$.

\subsection{Magnon}
Next, we derive the magnon dispersion of the FMI put on the surface of the TCI. 
We assume no lattice mismatch. 
The Hamiltonian of FMI is written as
\begin{equation}
    \begin{split}
        H_{\rm FMI} &= 
    -J_{1}\sum_{\ab<{i}, {j}>}
    \vv{S}_{{i}} \cdot \vv{S}_{{j}}
    -K\sum_{{i}}
    S^{z2}_{{i}},
    \end{split}
\end{equation}
where $\vv{S}_i = (S_{{i}}^x, S_{{i}}^y, S_{{i}}^z)$ denotes the $i$th spin located on $\boldsymbol{r}_i$, which spans the square lattice. 
The parameter $J_{1}>0$ is the ferromagnetic coupling constant between nearest-neighbor sites.
The parameter $K>0$ represents the easy-axis anisotropy. 
We introduce a Holstein-Primakoff transformation \cite{Holstein_Primakoff} for spin operators 
\begin{align}
    S_{{i}}^+
    &=\sqrt{2s}a_{{i}}^{\mathrm{A}/\mathrm{B}}, 
    \\
    S_{{i}}^-
    &=\sqrt{2s}a_{{i}}^{\mathrm{A}/\mathrm{B} \dag}, 
    \\
    S_{{i}}^z &= s-a_{{i}}^{\mathrm{A}/\mathrm{B}\dag} a_{{i}}^{\mathrm{A}/\mathrm{B}},
\end{align}
with the magnon operator $a_{{i}}^\tau$ satisfying $[a_{{i}}^\tau, a_{{j}}^{\tau'}] = 0$ and $[a_{{i}}^\tau, a_{{j}}^{\tau' \dag}] = \delta_{{ij}} \delta_{\tau\tau'}$.
Here, in ``A/B," A (B) applies when $\boldsymbol{r}_i$ belongs to sublattice A (B).
Performing the Fourier transformation $a_{{i}}^{\mathrm{A}/\mathrm{B}} = (1/\sqrt{
{N'}})\sum_{\vv{q}}a_{\vv{q}}^{\mathrm{A}/\mathrm{B}}e^{-i \boldsymbol{q} \cdot \boldsymbol{r}_i}$, the Hamiltonian of FMI is written as
\begin{align}
    &H_{\rm FMI}
    = \sum_{\vv{q}\in \rm{BZ}}a_{\vv{q}}^{\dag} \omega_{1\vv{q}} a_{\vv{q}}, 
    \\
    &\omega_{1\vv{q}} = \mqty( 4J_{1}s+2Ks & \gamma_{1\vv{q}} \\
                             \gamma_{1\vv{q}}                & 4J_{1}s+2Ks ), \label{Mat_FMI_magnon}
    \\
    &\gamma_{1\vv{q}} = 
    -4J_{1}s \cos\frac{q_x}{2} \cos\frac{q_y}{2},
\end{align}
where we define $a_{\vv{q}}^{\dag}=(a_{\vv{q}}^{A\dag}, a_{\vv{q}}^{B\dag})$, and wavevector $\vv{q}$ is defined in the two-dimensional BZ, $-\pi/a < q_x, q_y < \pi/a$. 
Diagonalizing Eq.~(\ref{Mat_FMI_magnon}), we get the dispersion of ferromagnetic magnon as
\begin{align}
    \omega_{\pm,\vv{q}} = 4J_{1}s+2Ks \pm \gamma_{1\vv{q}}.
\end{align}

\subsection{Wallpaper fermion--magnon interaction}
The isotropic exchange coupling between spins of the FMI and the wallpaper fermion is written in real space as follows;
\begin{align}
    H_{\rm int}
 = -\bar{J}
 \sum_{i}
 \sum_{\nu=x,y,z}
 \sum_{\sigma\sigma'}
 c_{i,0,\sigma}^{\dag}
 (\sigma^\nu)_{\sigma \sigma'}
 c_{i,0,\sigma'}
 \cdot
 S_{i}^{\nu},
\end{align}
where $c_{i,l,\sigma} \, (i,l \in \mathbb{Z},\, l\leq 0) $ is the annihilation operator of electron with spin $\sigma$ located at $(\boldsymbol{r}_i, a_z l)$.
Introducing the Fourier expansion of electron operator as $ c_{i, l} = (1/\sqrt{{N'}})\sum_{\vv{\kappa}\in \rm{BZ}} c_{\vv{\kappa},l}^{\mathrm{A}/\mathrm{B}} e^{-i\vv{\kappa} \cdot \vv{r}_{{i}} }$, we have
\begin{align}
		H_{\rm int} &=
		\sum_{\tau} \sum_{\vv{\kappa},\vv{q}\in \rm{BZ}}
		\Bigl[ 
		V \ab( a^{\tau}_{\vv{q}}
		c^{\tau\dag}_{\vv{\kappa}+\vv{q},0,\down}
		c^{\tau}_{\vv{\kappa},0,\up}+ \mathrm{h.c.})
	\notag\\&\hspace{6em}
		 -\bar{J}s \sum_{\sigma}\sigma c^{\tau\dag}_{\vv{\kappa},0,\sigma}c^{\tau}_{\vv{\kappa},0,\sigma} 
		 \Bigr],
\end{align}
up to the first order of $a$ and $a^\dag$, where $V=-\bar{J}\sqrt{2s}/\sqrt{{N'}}$.
In the last term of the above expression, coefficient $\sigma = \pm 1$ corresponds to spin up and down, respectively.

Next, we derive the low-energy effective interaction between the wallpaper fermions and magnons. 
As discussed in Sec.~\ref{sec:wf}, the electron is expanded by the wallpaper fermion as
\begin{align}
 c_{\boldsymbol{\bar M} + \boldsymbol{k}, 0, \sigma}^\tau
 = \eta_{\mathrm{surf}, 0} c_{\boldsymbol{k}, \sigma}^\tau,
\end{align}
where $\eta_{\text{surf}, l}$ is the wavefunction of the wallpaper fermion, which is the solution for the surface state of the three-dimensional Hamiltonian. 
Therefore, the Hamiltonian of interactions between the magnons and wallpaper fermions is written as
\begin{equation}
    \begin{split}
        H_{\rm int}
        =\sum_{\tau}\sum_{\vv{k},\vv{q}}  &\left[ 
        \tilde{V}
        \ab( a^{\tau}_{\vv{q}}c^{\tau\dag}_{\vv{k}+\vv{q},\down}c^{\tau}_{\vv{k},\up}+ \mathrm{h.c.}) \right.
        \\
        &\left. 
        -\tilde{J} s \sum_{\sigma}\sigma c^{\tau\dag}_{\vv{k},l,\sigma}c^{\tau}_{\vv{k},\sigma} \right],
    \end{split}
\end{equation}
with $\tilde{V} = V |\eta_{\mathrm{surf},0}|^2$ and $\tilde{J} = \bar J |\eta_{\mathrm{surf},0}|^2$.
The last term, induced Zeeman interaction, is regarded as the mass term of wallpaper fermions.

In summary, the Hamiltonian for the massive wallpaper fermion $H_{\mathrm{mwp}}$, ferromagnetic magnons $H_{\mathrm{FMI}}$, and the interaction between them $H_{\mathrm e-\mathrm{m}}$ are written as
\begin{align}
    H_{\rm mwp} &= \sum_{\vv{k}}c^{\dag}_{\vv{k}}(H_{\rm wp}(\vv{k})
    -\tilde{J}
    s\sigma_{z}\tau_{0})c_{\vv{k}}, \\
    H_{\rm FMI} &= \sum_{\vv{q}}a_{\vv{q}}^{\dag} \omega_{1\vv{q}} a_{\vv{q}}, \\
    H_{\rm e-m} &= 
    \tilde V
    \sum_{\tau}\sum_{\vv{k},\vv{q}}
    \ab( a^{\tau}_{\vv{q}}c^{\tau\dag}_{\vv{k}+\vv{q}\down}c^{\tau}_{\vv{k}\up}+ \mathrm{h.c.}), \\
    H &= H_{\rm mwp} + H_{\rm FMI} + H_{\rm e-m}.
\end{align}
In what follows, the matrix $\omega_{1\vv{q}}$ and the magnon dispersion $\omega_{\pm,\vv{q}}$ are approximated up to $q^2$ as
\begin{align}
{
 \omega_{+,\boldsymbol{q}}}
 &\approx
 2sK + \frac{sJ_1}{2} q^2,
 \label{w_mag_+}
 \\
{
 \omega _{-,\boldsymbol{q}}}
 &\approx
 8sJ_1 + 2sK - \frac{sJ_1}{2} q^2.
\end{align}

\subsection{Hamiltonians in the band basis}
We transform the basis of wallpaper fermions and magnons to the basis diagonalizing them as follows;
\begin{align}
    H_{\rm mwp} &= 
    \sum_{m=1}^4
    \sum_{\vv{k}}\xi_{m,\vv{k}}\psi_{m,\vv{k}}^{\dag}\psi_{m,\vv{k}},
    \\
    H_{\rm FMI} &= \sum_{\al=\pm}\sum_{\vv{q}}\omega_{\al,\vv{q}}{\tilde{a}_{\alpha, \boldsymbol{q}}^{\dag}}
    {\tilde{a}_{\alpha, \boldsymbol{q}}},
    \\
    {H_{\rm e-m} }
    &= \sum_{\vv{k},\vv{q}}
    \sum_{\alpha=\pm}
    \sum_{m,n=1}^4
      M_{\vv{k},\vv{q}}^{\al,m,n}{\tilde{a}_{\alpha, \boldsymbol{q}}}
      \psi_{m,\vv{k}+\vv{q}}^{\dag}\psi_{n,\vv{k}}
      + \mathrm{h.c.}, 
      \label{Hem}
    \\
    c^{\tau}_{\vv{k}\sigma} &= \sum_{m=1}^{4} Q^{\tau,\sigma}_{m,\vv{k}}\psi_{m,\vv{k}},
    \quad
    {
    a^{\tau}_{\vv{q}}} 
    {= \sum_{\alpha=\pm}
    w_{\alpha, \boldsymbol{q}}^\tau
    \tilde{a}_{\alpha, \boldsymbol{q}}},
    \\
    {
    w^A_{+,\boldsymbol{q}}} &= 
    w^A_{-,\boldsymbol{q}} = 
    w^B_{+,\boldsymbol{q}} = \frac{1}{\sqrt 2},
    \
    w^B_{-,\boldsymbol{q}} = -\frac{1}{\sqrt 2},
    \\
    M_{\vv{k},\vv{q}}^{\al,m,n} &= 
    {\tilde V}
    \sum_{\tau=A,B} 
    {w}_{\al,\vv{q}}^{\tau}Q_{m,\vv{k}+\vv{q}}^{\tau,\down*} Q_{n,\vv{k}}^{\tau,\up} \label{def_M}
    \\
    \xi_{m,\vv{k}} &= \pm \frac{1}{2}\sqrt{(v_{2}^{2}+v_{s1}^{2})k^2+4\tilde{J}^2 s^2 \mp 4v_{2}v_{s1}k_x k_y } - \mu,
\end{align}
where $v_2 = \sqrt{v_{s2}^2+v_{s2}^{\p 2}}$.
It is beneficial for the following calculations to derive the analytical representation of matrix $M$.
The matrix $Q$ is analytically given as
\begin{align}
    Q &=
    \frac{e^{-i\phi/2}}{\sqrt{2}}\mqty(
    u_{+}^{+} &  {u_{{-}}^{{+}}} & {u_{{+}}^{{-}}} & u_{-}^{-} 
    \\
    e^{i\phi}v_{+}^{+}  & {e^{i\phi}v_{{-}}^{{+}}} & {e^{i\phi}v_{{+}}^{{-}}} & e^{i\phi}v_{-}^{-} \\
    ie^{i\phi}u_{+}^{+} & {-ie^{i\phi}u_{{-}}^{{+}}} & {ie^{i\phi}u_{{+}}^{{-}}} & -ie^{i\phi}u_{-}^{-} \\
    -iv_{+}^{+} & {iv_{{-}}^{{+}}} & {-iv_{{+}}^{{-}}} & iv_{-}^{-}
    ),
\end{align}
with $\phi = \arg(v_{s2}+iv_{s2}')$.
\begin{align}
 Q^\dag
 \qty[H_{\rm wp}(\boldsymbol{k}) - s \tilde J \tau_0 \sigma_z]
 Q
 = \mathrm{diag}(\xi_{1\boldsymbol{k}}, \xi_{2\boldsymbol{k}}, \xi_{3\boldsymbol{k}}, \xi_{4\boldsymbol{k}}),
\end{align}
\begin{align}
 \xi_{1\boldsymbol{k}} &= \frac{1}{2}\sqrt{(v_{2}^{2}+v_{s1}^{2})k^2+4\tilde{J}^2 s^2 - 4v_{2}v_{s1}k_x k_y } - \mu,
 \\
 \xi_{2\boldsymbol{k}} &= 
 {
 \frac{1}{2}\sqrt{(v_{2}^{2}+v_{s1}^{2})k^2+4\tilde{J}^2 s^2 + 4v_{2}v_{s1}k_x k_y } - \mu},
 \\
 \xi_{3\boldsymbol{k}} &=
 {-\xi_{1\boldsymbol{k}}},
  \\
  \xi_{4\boldsymbol{k}} &= 
  {- \xi_{2\boldsymbol{k}}}.
\end{align}
\begin{align}
    \mqty(
    u_{\pm}^{{\eta}} \\ v_{\pm}^{{\eta}}
    )
    &=
    \mqty(
    \sqrt{\displaystyle\frac{1}{2}\lb 1 {+\eta} \frac{{-s\tilde J}}{{|\xi_{\pm}+\mu}|} \rb} \\
    {\eta}
    e^{i\varphi_{\pm}}
    \sqrt{\displaystyle\frac{1}{2}\lb 1 {-\eta} 
    	\frac{{-s\tilde J}}{{|\xi_{\pm}+\mu|}} \rb}
    ), 
    \
    \eta = \pm,
    \\
    \varphi_{\pm} &= \mathrm{arg} \LB v_{2}(k_x+ik_y) \mp iv_{s1}(k_x-ik_y) \RB. \label{def_varphi}
\end{align}
In this representation, we can write the matrix element of $M$ as
\begin{align}
    M^{\pm,1,1}_{\vv{k},\vv{q}} &= 
    \frac{{\tilde V}}{2\sqrt{2}} \lb e^{-i\phi} \mp e^{i\phi} \rb v_{+}^{+*}(\vv{k}+\vv{q})u_{+}^{+}(\vv{k}), 
    \label{eq:M1}
    \\
    M^{\pm,1,2}_{\vv{k},\vv{q}} &= 
    \frac{{\tilde V}}{2\sqrt{2}} \lb e^{-i\phi} \pm e^{i\phi} \rb v_{+}^{+*}(\vv{k}+\vv{q})
    u_{{-}}^{{+}}(\vv{k}), 
    \label{eq:M2}
    \\
    M^{\pm,2,1}_{\vv{k},\vv{q}} &= 
    \frac{{\tilde V}}{2\sqrt{2}} \lb e^{-i\phi} \pm e^{i\phi} \rb v_{-}^{+*}(\vv{k}+\vv{q})u_{+}^{+}(\vv{k}), 
    \label{eq:M3}
    \\
    M^{\pm,2,2}_{\vv{k},\vv{q}} &= 
    \frac{{\tilde V}}{2\sqrt{2}} \lb e^{-i\phi} \mp e^{i\phi} \rb v_{-}^{+*}(\vv{k}+\vv{q})
    u_{{-}}^{{+}}(\vv{k}),
    \label{eq:M4}
\end{align}
In the above equations and the following analysis, we focus exclusively on the two conduction bands $\xi_{1\boldsymbol{k}}$ and $\xi_{2\boldsymbol{k}}$ that cross the Fermi energy.

\section{BCS pairing}\label{sec_BCS}
The superconducting transition temperature is determined by the linearized Eliashberg equation.
In this study, we adopt no retardation effect. 
The detail of the derivation is shown in Appendix \ref{app:Eliashberg}.
As a result, we can write the linearized Eliashberg equation for the BCS pairing with no center-of-mass momentum as
\begin{align}
    \Delta_{n,n^\p}(\vv{k})
    &=-\sum_{m,m^\p,\vv{k}\p}
    V^{m,m^\p;n,n^\p}_{\vv{k},\vv{k}'} 
    \chi^{m,m^\p}_{\vv{k^\p}}
        \Delta_{m,m^\p}(\vv{k}^\p),\label{gap_BCS}
    \\
    \chi^{m,m^\p}_{\vv{k^\p}}
    &= \frac{ 1 - f(\xi_{m,\vv{k}^\p}) - f(\xi_{m^\p,\vv{k}^\p}) }{ \xi_{m,\vv{k}^\p}+\xi_{m^\p,\vv{k}^\p} },
    \label{def_sus_BCS}
\end{align}
where $\chi^{m,m^\p}_{\vv{k^\p}}$ is the superconducting susceptibility of wallpaper fermions and $f(\xi_{m,\vv{k}^\p}) =(e^{\xi_{m, \boldsymbol{k}'}/T} + 1)^{-1}$ is the Fermi-Dirac distribution function.
The effective interaction is given by (see Appendix \ref{app:Eliashberg})
\begin{align}
 V_{\boldsymbol{k}, \boldsymbol{k'}}^{m,m'; n,n'}
 &= 
  -\sum_{\alpha} 
 \frac{
 	M_{\boldsymbol{k}', \boldsymbol{k}-\boldsymbol{k}'}^{\alpha, n, m}
 	M_{-\boldsymbol{k}, \boldsymbol{k}-\boldsymbol{k}'}^{*\alpha, m', n'}
 	}
 	{\omega_{\alpha, \boldsymbol{k}-\boldsymbol{k}'}}
 \notag\\&\quad
 -
 \sum_{\alpha}
 \frac{
 	M_{-\boldsymbol{k}', \boldsymbol{k}'-\boldsymbol{k}}^{\alpha, n', m'}
 	M_{\boldsymbol{k}, \boldsymbol{k}'-\boldsymbol{k}}^{*\alpha, m, n}}
 	{\omega_{\alpha, \boldsymbol{k}-\boldsymbol{k}'}}.
\end{align}

\subsection{Solution to the linearized Eliashberg equation in the isotropic limit}
This section determines the symmetry of the superconducting state by solving the linearized Eliashberg equation (\ref{gap_BCS}).
As shown below, in the isotropic limit of $v_{s1}=0$, the linearized Eliashberg equation simplifies and demonstrates that the chiral $p$-wave state is the most stable.

In the isotropic limit of $v_{s1}=0$, 
\begin{align}
    \varphi_{+}(\boldsymbol{k})
    &=
    \varphi_{-}(\boldsymbol{k})
    =
    \varphi_{\boldsymbol{k}},
\end{align}
where $\varphi_{\boldsymbol{k}}$ denotes the azimuth angle of $\boldsymbol{k}$. 
The effective interaction reduces to
\begin{align}
    V_{\boldsymbol{k}', \boldsymbol{k}}^{m,m'; n,n'}
    &=
    \frac{1}{32}
    \ab(1-\frac{s^2\tilde J^2}{\mu^2})
    \frac{\tilde V^2}{\omega_{+, \boldsymbol{k}-\boldsymbol{k}'}}
    e^{i \varphi_{\boldsymbol{k}'}}
    e^{-i \varphi_{\boldsymbol{k}}}
    \notag\\&\quad\times
    \ab(1+mm'nn')
    Y_{mn,m'n'}^*,
    \\
    Y_{mn, m'n'}^*
    &= 
    \ab(e^{i\phi} - mn e^{-i \phi})
    \ab(e^{-i\phi}-m'n' e^{i\phi}),
\end{align}
where we assume $\omega_{+, \boldsymbol{q}} \ll \omega_{-, \boldsymbol{q}}$. 
The susceptibility also becomes isotropic, $\chi_{\boldsymbol{k}}^{m,m'} \equiv \chi_{k}$.

The angular-momentum $l$ solution has the form
\begin{align}
 \Delta_{n,n'}(\boldsymbol{k})
 = 
 \Delta_{n,n'}^l(k)
 e^{i l \varphi_{\boldsymbol{k}}}.
\end{align}
The Fermi statistics lead to
\begin{align}
 \Delta_{n,n'}^l(k)
 = 
 (-1)^{l+1}
 \Delta_{n',n}^l(k).
 \label{fermi}
\end{align}
The Eliashberg equation is given by
\begin{align}
 \Delta_{n,n'}^l(k)
 &= 
 -A
 \sum_{mm'}
 \delta_{1,mm'nn'}
 Y_{mn,m'n'}^*
 \notag\\& \quad \times
 \int \frac{dk'}{2\pi} k'
 \frac{\tilde J^2 \chi_{k'}}{\omega_l(k,k')}
 \Delta_{m,m'}^l(k').
\end{align}
with the positive constant
\begin{align}
 A =  \frac{s}{8} 
 \ab(1-\frac{s^2\tilde{J}^2}{\mu^2}) > 0,
\end{align}
and 
\begin{align}
 \frac{1}{\omega_l(k,k')}
 = 
 \int \frac{d\varphi_{\boldsymbol{k}'}}{2\pi}
 \frac{
 	e^{-i (l+1) (\varphi_{\boldsymbol{k}} - \varphi_{\boldsymbol{k'}})}
 }
 {\omega_{+,\boldsymbol{k}-\boldsymbol{k'}}}.
\end{align}
The magnon dispersion relation for the long-wavelength limit is given by
\begin{align}
 \omega_{+,\boldsymbol{k}-\boldsymbol{k'}}
 &
 = 2sK + \frac{sJ_1}{2}
 |\boldsymbol{k}-\boldsymbol{k'}|^2
 \notag\\&
 = 
 2sK + \frac{sJ_1}{2} 
 \ab[k^2+k'^2 -2 kk' \cos(\varphi_{\boldsymbol{k}}-\varphi_{\boldsymbol{k'}})],
\end{align}
which is also a function of $\varphi_{\boldsymbol{k}}-\varphi_{\boldsymbol{k}'}$.
Thus we find
\begin{align}
 \frac{1}{\omega_l(k,k')}
  = \int \frac{d(\varphi_{\boldsymbol{k}}-\varphi_{\boldsymbol{k'}})}{2\pi}
  \frac{e^{-i (l+1) (\varphi_{\boldsymbol{k}} - \varphi_{\boldsymbol{k'}})}}{\omega_{+, \boldsymbol{k}-\boldsymbol{k}'}}.
\end{align}
Since $\omega_{+, \boldsymbol{k}-\boldsymbol{k'}}$ is an even function of $\varphi_{\boldsymbol{k}}-\varphi_{\boldsymbol{k'}}$, we have
\begin{align}
 \frac{1}{\omega_l(k,k')}
 = \int_{0}^\pi \frac{d(\varphi_{\boldsymbol{k}}-\varphi_{\boldsymbol{k'}})}{\pi}
 \frac{\cos\ab((l+1) (\varphi_{\boldsymbol{k}'} - \varphi_{\boldsymbol{k}}))}
 {\omega_{+, \boldsymbol{k}-\boldsymbol{k}'}}.
\end{align}
The above integral takes a positive value for any $l$, because $\omega_{+, \boldsymbol{k}-\boldsymbol{k'}}^{-1}$ is a monotonically decreasing function in $\varphi_{\boldsymbol{k}}-\varphi_{\boldsymbol{k}'} \in [0, \pi]$. 

For an even $l$, $\Delta_{11}^l(k) = \Delta_{22}^l(k) = 0$, due to the Fermi statistics Eq.~(\ref{fermi}).
For the off-diagonal component, $\Delta_{12}^l(k) = - \Delta_{21}^l(k)$, the Eliashberg equation is given by
\begin{align}
 \Delta_{12}^l(k)
 = 4A
 \cos2\phi
 \int \frac{dk'}{2\pi} k' \frac{\tilde J^2 \chi_{k'}}{\omega_l(k,k')}
 \Delta_{12}^l(k').
\end{align}
All the coefficients in the above expression are positive for $\cos2\phi>0$, in which a nontrivial solution could be obtained. 
In the weak-coupling limit, the Eliashberg equation is simplified on the Fermi surface as
\begin{align}
 1
 = 4A \cos2\phi
 \frac{\tilde J^2 \chi(T_{\mathrm{c}})}{\omega_l(k_{\mathrm{F}}, k_{\mathrm{F}})},
\end{align}
and the transition temperature $T_{\mathrm{c}}$ is obtained from
\begin{align}
 \chi(T_{\mathrm{c}})
 \simeq D(\epsilon_{\mathrm{F}})
 \int_{-\epsilon_{\mathrm{c}}}^{\epsilon_{\mathrm{c}}} 
 d\xi_{k'}
 \chi_{k'}
 \simeq 
 D(\epsilon_{\mathrm{F}})
 \ln\frac{1.14 \epsilon_{\mathrm{c}}}{T_{\mathrm{c}}},
\end{align}
where $D(\epsilon_{\mathrm{F}})$ denotes the density of states on the Fermi energy. 

On the other hand, for an odd $l$, $\Delta_{n,n'}^l(k)=\Delta_{n',n}^l(k)$, the Eliashberg equation is written as
\begin{align}
	\begin{pmatrix}
		\Delta_{11}^l(k)
		\\
		\Delta_{22}^l(k)
	\end{pmatrix}
	=
	-A
	\int \frac{dk'}{2\pi}
	k' \frac{\tilde J^2 \chi_{{k'}}}{\omega_l(k,k')}
	\Phi
		\begin{pmatrix}
		\Delta_{11}^l(k')
		\\
		\Delta_{22}^l(k')
	\end{pmatrix},
\end{align}
with the matrix
\begin{align}
	\Phi
	=
		\begin{pmatrix}
		2(1-\cos2\phi) & 2(1+\cos2\phi)
		\\
		2(1+\cos2\phi) & 2(1-\cos2\phi)
	\end{pmatrix}.
\end{align}
The eigenvalues of $\Phi$ are given by $4$ and $-4 \cos2\phi$.
There could be a nontrivial solution for $\cos2\phi>0$. 
The off-diagonal component satisfies the following Eliashberg equation
\begin{align}
 \Delta_{12}^l(k)
 = 
 -4A
 \int \frac{dk'}{2\pi}
 k' \frac{\tilde J^2 \chi_{{k'}}}{\omega_l(k,k')}
 \Delta_{12}^l(k'),
 \label{gap_eq_BCS_sym}
\end{align}
which could have no nontrivial solution. 

The positive eigenvalues of the Eliashberg equations of $\Delta_{1,2}^{l}$ and $(\Delta_{1,1}^{l}, \Delta_{2,2}^{l})$ is given by
\begin{align}
 4A \cos2\phi \frac{\tilde J^2 \chi(T_{\mathrm{c}})}{\omega_l(k_{\mathrm{F}}, k_{\mathrm{F}})}.
 \label{eq:eigenvalue}
\end{align}
Therefore, the pair potential is most stable for the smallest $\omega_l(k_{\mathrm{F}}, k_{\mathrm{F}})$, the explicit expression of which is derived by the contour integral
\begin{align}
 \frac{1}{\omega_{l}(k,k')}
 &= \frac{1}{C} \mathrm{Re} 
 \oint_{|z|=1} \frac{dz}{2\pi iz} 
 \frac{z^{l+1}}{1-B (z+z^{-1})/2}
 \notag\\
 &=
 \frac{1}{C} 
 \frac{\alpha^{l+1}}{\sqrt{1-B^2}},
 \\
 \alpha &=
 \frac{1}{B} - \sqrt{\frac{1}{B^2}-1},
 \\
 C &=
 2sK + \frac{sJ_1}{2} \ab(k^2+k'^2),
 \\
 B &=
 \frac{sJ_1kk'}{2sK + ({sJ_1}/{2}) \ab(k^2+k'^2)},
\end{align}
for $l \geq -1$.
Note that $|B| < 1$ and $|\alpha| < 1$, hence $1/\omega_{l}(k,k')$ takes the maximum value for $l=-1$. 
Consequently, the $l=-1$ (chiral $p$-wave) solution is most stable when $\cos2\phi > 0$.

\subsection{Numerical analysis}
Now, we evaluate the critical temperature by solving Eq.~(\ref{gap_eq_BCS_sym}) for $l = -1$, which is the most stable. 
We introduce a discretization of the integral of a function $g( \vv{k})$ that takes a larger value near the Fermi surface, $k_1 < k < k_2$, as
\begin{align}
    \frac{1}{N'}
    \sum_{\vv{k}\in \rm{BZ}} g(\vv{k})
    &\simeq \int_{\rm BZ}\frac{d^2 k}{(2\pi)^2}g(\vv{k})
    \notag\\
    &\simeq \int_{k_1}^{k_2}\frac{dk}{2\pi} k\, g'(k)
    \notag \\
    &\simeq \frac{1}{2\pi} \sum_{k}\frac{k_2-k_1}{\sqrt{N}}k\, g'(k)
    \notag\\
    &=\frac{k_2-k_1}{2\pi \sqrt{N}}\sum_{k}k\, g'(k),
\end{align}
where 
\begin{align}
    g'(k) = \int_{0}^{2\pi} \frac{d\theta}{2\pi} g(\vv{k}).
\end{align}

\begin{figure}
    \centering
    \subfigure[]{
    \centering
    \includegraphics[scale=0.66]{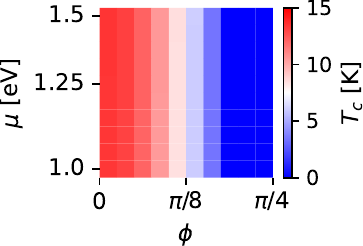}
    \label{TC_BCS_mu_phi}
    }
    \subfigure[]{
    \centering
    \includegraphics[scale=0.66]{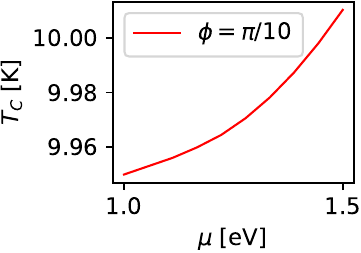}
    \label{TC_BCS_mu}
    }
    \caption{Critical temperature $T_{\mathrm{c}}$ of the BCS pairing by the Eliashberg equation (\ref{gap_eq_BCS_sym}).
    The parameters are set as $v_2 = 0.06$ eV, $v_{s1} = 0.0$ eV, $\bar{J} = 10.0$ meV, $J_1 = K \times 10^{4} = 10.0$ mev, $\sqrt{N} = 200$, $k_1=0.95k_{\rm F}$, and $k_2 = 1.05 k_{\rm F}$, where $k_{\rm F}$ is the Fermi wavevector.
    (a) $T_{\mathrm{c}}$ becomes lower with increasing $\phi$.
    (b) $T_{\mathrm{c}}$ increases with the chemical potential $\mu$.}
    \label{fig:TC-BCS}
\end{figure}

In Fig.~\ref{fig:TC-BCS}, we show the critical temperature $T_{\mathrm{c}}$ calculated by the Eliashberg equation (\ref{gap_eq_BCS_sym}) with the above discretization.
From Fig.~\ref{TC_BCS_mu_phi}, we find that the $T_{\mathrm c}$ decrease as $\phi \in [0,\pi/4]$ increases.
This is consistent that the eigenvalue of the linearized Eliashberg equation is proportional to $\cos2\phi$ [Eq.~(\ref{eq:eigenvalue})]. 
Additionally, Fig.~\ref{TC_BCS_mu} shows that the $T_{\mathrm{c}}$ increases with the chemical potential $\mu$.
This is understood as a consequence of the density of states increasing with the chemical potential.
In conclusion, the chiral $p$-wave superconducting state, $l=-1$, is realized with the BCS pairing.

\section{Amperean pairing}\label{sec_Amp}
In this section, we consider the Amperean pairing, in which electrons with momenta $\boldsymbol K + \boldsymbol{p}$ and $\boldsymbol{K}-\boldsymbol{p}$ form a Cooper pair, which has the center-of-mass momentum $2\vv{K}$.
The linearized Eliashberg equation for the Amperean pairing as (see Appendix~\ref{app:Eliashberg})
\begin{align}
&
    \Delta_{n,n'}(\vv{p})
    =-
      \sum_{\vv{p}'}
      U_{O,\vv{p},\vv{p}',\vv{K}}^{m,m';n,n'}
      \chi^{m,m^\p}_{\vv{p}^\p,\vv{K}} 
      \Delta_{m,m'}(\vv{p}')
      \label{Simp_gap_eq},
      \\
     &
      U^{m,m^\p;n,n^\p}_{O,\vv{p},\vv{p}^\p,\vv{K}}
 =
      \frac{
      	U^{m,m^\p;n,n^\p}_{\vv{p},\vv{p}^\p,\vv{K}}
      	- 
      	U^{m^\p,m;n,n^\p}_{\vv{p},-\vv{p}^\p,\vv{K}}
      	}
      {2},\label{V_odd_Amp}
      \\
  	&
  	U_{\boldsymbol{p}, \boldsymbol{p}', \boldsymbol{K}}^{m,m';n,n'}
  	\notag\\&
  	= 
  	-\sum_{\alpha} 
  	\frac{
  		M_{\boldsymbol{K}+\boldsymbol{p}', \boldsymbol{p}-\boldsymbol{p}'}^{\alpha, n, m}
  		M_{\boldsymbol{K}-\boldsymbol{p}, \boldsymbol{p}-\boldsymbol{p}'}^{*\alpha, m', n'}
  	}
  	{\omega_{\alpha, \boldsymbol{p}-\boldsymbol{p}'}}
  	\notag\\&\quad
  	-
  	\sum_{\alpha}
  	\frac{
  		M_{\boldsymbol{K}-\boldsymbol{p}', \boldsymbol{p}'-\boldsymbol{p}}^{\alpha, n', m'}
  		M_{\boldsymbol{K}+\boldsymbol{p}, \boldsymbol{p}'-\boldsymbol{p}}^{*\alpha, m, n}}
  	{\omega_{\alpha, \boldsymbol{p}-\boldsymbol{p}'}},
  \\
      &\chi^{m,m^\p}_{\vv{p}^\p,\vv{K}}
 = \frac{1-f(\xi_{m,\vv{K}+\vv{p}}) - f(\xi_{m',\vv{K}-\vv{p}}) }{\xi_{m,\vv{K}+\vv{p}} + \xi_{m',\vv{K}-\vv{p}} }.
 \label{def_sus_Amp}
\end{align}

\subsection{Isotropic limit}\label{sec:isotropic}

In the following, we clarify the behavior of the attraction in the isotropic limit and discuss the symmetry of potential solutions to the gap equation.
In the vicinity of the Fermi surface, $\boldsymbol{p}, \boldsymbol{p'} \sim \boldsymbol{0}$, in the isotropic case, $v_{s1}=0$, the energy denominators are approximated as
\begin{align}
 \frac{1}{\omega_{+, \boldsymbol{p}-\boldsymbol{p}'}}
 \simeq
 -\ab(
 \frac{1}{2sK}
 +
 \frac{J_1}{4sK^2} \boldsymbol{p} \cdot \boldsymbol{p}'
 ).
 \label{eq:denom}
\end{align}
Here we assume $\omega_{+, \boldsymbol{p}-\boldsymbol{p}'} \ll \omega_{-, \boldsymbol{p}-\boldsymbol{p}'}$.
The electron-magnon coupling constant is also approximated as $M_{\boldsymbol{K}-\boldsymbol{p}', \boldsymbol{p}'-\boldsymbol{p}}^{\beta, n', m'} \simeq M_{\boldsymbol{K}, \boldsymbol{0}}^{\beta, n', m'}$.
Therefore, we find
\begin{widetext}
\begin{align}
	U^{m,m';n,n'}_{O,\vv{p},\vv{p}',\vv{K}}
	\simeq
	-\frac{\tilde{V}^2 X_{\vv{K}}}{2}\times
	\begin{cases}
		\displaystyle\frac{J_{1}\cos^2 \phi}{2sK^2}\vv{p}\cdot\vv{p}'
		,
		& m=m'=n=n',
		\\[1em]
		\displaystyle\frac{J_{1}\sin^2 \phi}{2sK^2}\vv{p}\cdot\vv{p}'
		,
		& m=m'=-n=-n',
		\\[1em]
		\displaystyle\frac{\cos 2\phi}{2sK} 
		+ \frac{J_{1}\vv{p}\cdot\vv{p}'}{4sK^2},
		& m=-m'=n=-n',
		\\[1em]
		-\displaystyle\frac{\cos 2\phi}{2sK} 
		+ \frac{J_{1}\vv{p}\cdot\vv{p}'}{4sK^2},
		& m=-m'=-n=n',
		\\[1em]
		0, & \text{otherwise}.
	\end{cases}
\label{eq:isotropic-limit}
\end{align}
\end{widetext}
with
\begin{align}
		X_{\boldsymbol{K}}
		= \ab|
		v_{+}^+(\boldsymbol{K})
		u_{+}^+(\boldsymbol{K})
		v_{+}^+(\boldsymbol{K})
		v_{+}^+(\boldsymbol{K})
		|.
\end{align}
The detail of the derivations is shown in Appendix \ref{App_isotoropic}.
In the former two cases, $m=m'$ and $n=n'$, for the intraband pairings, the interaction is proportional to $\boldsymbol{p} \cdot \boldsymbol{p}'$, which leads to the $p$-wave solution to the gap equation.
On the other hand, in the latter two cases, $m=-m'$ and $n=-n'$, for the interband pairings, the interaction can induce the parity mixing of the $s$-wave and $p$-wave pairings. 
Therefore, a parity-mixing superconducting state is probable when the easy-axis anisotropy $K$ is larger than the exchange coupling $J_1$.

\subsection{Numerical analysis}
\begin{figure}
	\includegraphics[scale=1.0]{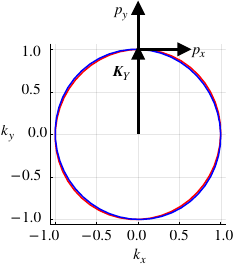}
	\label{Fermi_surf_Amp}
	\caption{
	Fermi surface of the wallpaper fermion. 
	$\vv{K}_Y$ is the center-of-mass momentum of the Amperean pair and $(p_x, p_y)$ is the local coordinate.
	The parameters are taken as $\tilde{J}=10.0$ meV, $s=1/2$ , $\mu = 0.6$ eV, $v_2=1.2$ eV, $v_{s1}=0$ for the perfect circle (red), and $v_{s1}= 1.0$ meV for the slightly deformed circle (blue). 
 }
	\label{fig:FermiSurface}
\end{figure}

\begin{figure}
	\centering
	\includegraphics[scale=1.1]{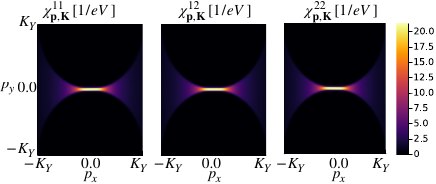}
	\caption{
		The susceptibility for the Amperean pairing for $\vv{K}_{Y}=K_Y \hat{\boldsymbol{y}}$.
		We set the parameters as 
		$\tilde J = 10.0$ meV, $s=1/2$, $\mu=0.6$ eV,
		$v_2=1.2$ eV, $v_{s1}= 2.0$ meV, and $T=100$ K.
	}%
	\label{fig_sus_Amp}
\end{figure}

The above analysis is verified by the numerical solution of the gap equation (\ref{Simp_gap_eq}) in the same manner as used in Ref.~\cite{MagnonSC_Erlandsen}. 
In the isotropic limit $v_{s1}=0$, the energy bands are doubly degenerate, and the Fermi surface is the circle, as shown in Fig.~\ref{fig:FermiSurface}.
The susceptibility for the Amperean pairing is defined in Eq.~(\ref{def_sus_Amp}) and shown in Fig.~\ref{fig_sus_Amp} for the center-of-mass momentum $\boldsymbol{K} = \vv{K}_Y := K_Y \hat{\boldsymbol{y}}$ located on the Fermi surface, $\xi_{m, \boldsymbol{K}_Y} = 0$.
The summation over $\boldsymbol{p}'$ in the gap equation is restricted to the region where the susceptibility is large as
\begin{align}
 \sum_{\boldsymbol{p}'}
 |p_{x}'| \leq p_{\rm c}, 
 \
 |p_{y}'| \leq p_{x}^{\prime 2}/p_{\rm c},
\end{align}
Now we define the new variables as $p_x'=p_1$, $p_y' = p_2 \cdot p_{1}^2/p_{\mathrm{c}}$, within $|p_1| \leq p_{\rm c}$ and $|p_2| \leq 1$.
We set $p_{\mathrm c} = K_Y/10$. 
The Jacobian is $|J| = p_{1}^2/p_{\mathrm{c}}$.
Therefore, the sum of the wavevector is evaluated as follows;
\begin{align}
	\frac{1}{N'}\sum_{\vv{p}'} g(\boldsymbol{p}')
	&\simeq \int \frac{d^2p'}{(2\pi)^2}g(\vv{p}')
	\notag\\
	&\simeq \frac{1}{(2\pi)^2}\int_{-p_c}^{p_c}dp_1 \int_{-1}^{1}dp_2 \frac{p_{1}^2}{p_c}g(\vv{p}')
	\notag\\
	&\simeq \frac{1}{p_c(2\pi)^2} \sum_{p_1,p_2}p_{1}^2\frac{2p_{c}}{\sqrt{N}-1} \frac{2}{\sqrt{N}-1}g(\vv{p}')
	\notag \\
	&= \frac{1}{(\sqrt{N}-1)^2\pi^2}\sum_{p_1,p_2} p_{1}^2 g(\vv{p}'),
\end{align}
for the discretized points $p_1 = -p_{\rm c} + (2p_{\rm c}/(\sqrt{N}-1)) i$ and $p_2 = -1 + (2/(\sqrt{N}-1)) j$ with $i, j = 0, \cdots, \sqrt{N}-1$.

\begin{figure}
     \includegraphics[scale=0.6]{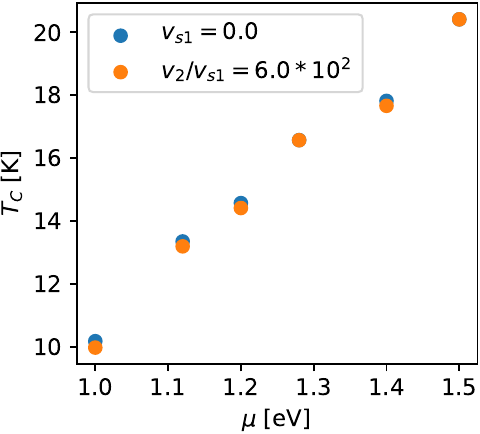}
     \caption{
	 Superconducting critical temperature $T_{\rm c}$ as a function of the chemical potential $\mu$.
	 The parameters are taken as $J_1 = 0.4$ eV and $K = 0.01$ meV 
	 The other parameters are taken as in Fig.~\ref{fig_sus_Amp}.
	}
 \label{Fig:Tc}
\end{figure}

\begin{figure*}
\begin{center}
    \includegraphics[width=1.0\linewidth]{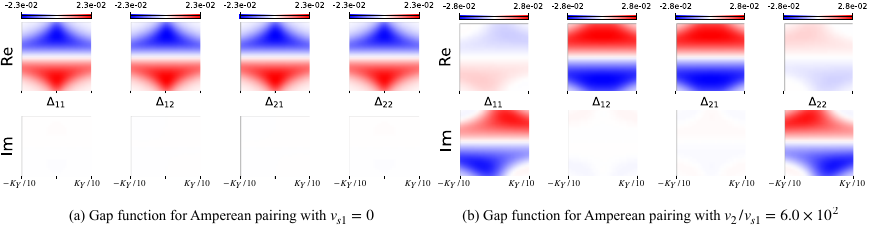}
    \caption{The numerical solutions of the gap equation Eq.~(\ref{Simp_gap_eq}).
    The horizontal and vertical axes denote $p_1$ and $p_2$, respectively.
    (a) The parameters are taken as 
    $\bar{J}=10.0$ meV, $N=900$, $J_1=0.4$ eV, $K = 0.01$ meV, $\phi=\pi/5$, $v_2=1.2$ eV, $v_{s1}= 0.0$ eV, and $\mu=0.9$ eV.
    (b) The parameters are the same as in (a) except for $v_2/v_{s1}=6.0\times 10^{2}$.
    }
    \label{fig_Amp_Y}
\end{center}
\end{figure*}

We show the critical temperature $T_{\rm c}$ as a function of the chemical potential $\mu$ in Fig.~\ref{Fig:Tc} for $v_{s1}=0$ and $v_{s2}/v_{s1}=600$.
The critical temperature is proportional to $\mu$.
In the slightly anisotropic case, $v_{2}/v_{s1} = 600$, we cannot find the solution with a finite critical temperature for $\mu < 1.0$ eV.
This result implies that the band anisotropy works against the superconducting transition.

Figure~\ref{fig_Amp_Y} shows the structure of the gap function as a function of $p_1$ and $p_2$ in the isotropic limit $v_{s1} \simeq 0$.
The gap symmetry is found to be the $p$-wave.
Conversely, when the easy-axis anisotropy $K$ is strong, the solution exhibits a substantial mixing of $s$-wave and $p$-wave pairings.
From Eq.~(\ref{eq:isotropic-limit}), we find that the momentum-independent terms inducing the $s$-wave pairing are proportional to $K^{-1}$ while the momentum-dependent terms inducing the $p$-wave pairing are proportional to $K^{-2}$.
\begin{figure}
    \centering
    \includegraphics[scale=1.1]{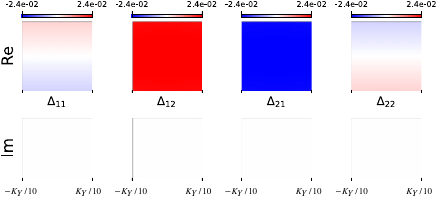}
    \caption{
    	The parity mixing solution of gap equation (\ref{Simp_gap_eq}) with parameters $\bar{J}=10.0$ meV, $N=900$, $J_1=0.4$ eV, $ K = 0.01$ eV, $\phi=\pi/5$, $v_2=1.2$ eV, $v_{s1}= 0.1$ eV, and $\mu=0.6$ eV.
    }
    \label{fig_gap_mix}
\end{figure}
Figure~\ref{fig_gap_mix} shows the parity mixing solution with maximum eigenvalue $\lambda \sim 10^{-2}$ for the strong easy-axis anisotropy $K=0.01$ eV. 
The intraband ($\Delta_{1,1}$ and $\Delta_{2,2}$) and interband ($\Delta_{1,2}$ and $\Delta_{2,1}$) components are dominated by $p$-wave and $s$-wave pairings, respectively. 

Contrary to the BCS pairing discussed in Sec.~\ref{sec_BCS}, parity mixing can occur in the Amperean pairing. 
This is because the Amperean pairing is free from the crystalline symmetry. 
The wallpaper fermions do not have a four-fold rotational symmetry around the center-of-mass momentum $\boldsymbol{K}$. 
Additionally, due to the mass term coming from the ferromagnetic moment, the reflection symmetry with respect to the $(100)$ plane is broken.
Therefore, it is permissible to mix the odd- and even-parity components in the gap function.

\section{Phase diagram of BCS and Amperean pairings}\label{sec_Amp_BCS}
\begin{figure}
    \centering
    \includegraphics[scale=0.9]{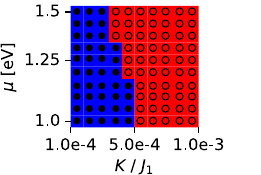}
    \caption{
    Phase diagram of BCS pairing vs. Amperean pairing with the isotropic band limit ($v_{s1}=0$).
    The parameters are set as $\phi = \pi/10$, $J_1  = 1.0$ meV, $\bar{J} = 10$ meV, and $v_2 = 0.6$ eV.
    The closed circles indicate (in the blue region) that Amperean pairing is stable, while the open circles (in the red region) indicate that BCS pairing is stable.
    }
    \label{fig:phase}
\end{figure}

This section discusses the stability of the BCS and Amperean pairings and determines the phase diagram.
To do this, we compare the critical temperature calculated by the linearized Eliashberg equations Eqs.~(\ref{gap_eq_BCS_sym}) and (\ref{Simp_gap_eq}).
In Fig.~\ref{fig:phase}, we show the phase diagram as a function of the chemical potential $\mu$ and the easy-axis parameter $K$.
The BCS pairing with chiral $p$-wave state is dominant in the region of high $K$.
On the other hand, the Amperean pairing state with the $p$-wave structure is realized for small $K$.
This result suggests that we can control the types of pairings by the external magnetic field or thickness of the ferromagnetic insulators.

\section{Conclusion}\label{sec_Cond}
We have investigated the two-dimensional superconductivity of wallpaper fermions mediated by ferromagnetic magnons at the interface between a nonsymmorphic TCI and FMI.
In the case of the isotropic band structure, superconducting instability toward the BCS pairing, which has no net momentum, is enhanced, leading to chiral $p$-wave superconductivity.
On the other hand, the Amperean pairing with finite center-of-mass momentum can emerge as a parity-mixed superconducting state.
In the isotropic limit, we demonstrated that the Amperean state exhibits the $p$-wave symmetry.
Furthermore, we discussed the effects of the band anisotropy of wallpaper fermion and the easy-axis anisotropy of FMI, which induce a solution by mixing $p$-wave and $s$-wave states.
Such mixing of superconducting symmetry is absent in the case of a single Dirac fermion on the surface of three-dimensional TIs.
By comparing the critical temperatures of the BCS and Amperean pairing states, we found that magnetization anisotropy plays a crucial role in determining the superconducting phase.
When the magnetization anisotropy is weak, BCS pairing is more stable than Amperean pairing, whereas the latter becomes stable for the strong anisotropy.
These results suggest that the type of superconducting pairing can be controlled by tuning external magnetic fields or the thickness of the FMI.

\begin{acknowledgments}
 This work is supported by JSPS KAKENHI for Grants (Grants Nos.~JP20K03835 and JP24H00853) and JST SPRING (Grant No. JPMJSP2125). 
\end{acknowledgments}

\appendix

\section{Derivation of effective action for fermions}\label{Driv_int_Eli}
\label{app:eff}
In this Appendix, we derive the effective action for fermions by integrating the magnons.
The action for the wallpaper fermion is given by
\begin{align}
    S_{\rm WPF}
    &= \int d\tau 
    \sum_{\boldsymbol{k}}
    {\sum_{m}}
    \bar{\psi}_{m,\vv{k}}(\partial_{\tau}+\xi_{m,\vv{k}})\psi_{m,\vv{k}},
\end{align}
where $\psi_{m,\vv{k}}$ and $\bar\psi_{m, \boldsymbol{k}}$ are Grassmann numbers.
The action of magnons is written as
\begin{align}
    S_{\rm FMI} = \int d\tau 
    \sum_{\boldsymbol{q}}
    {\sum_{\alpha}}
    \bar{a}_{\alpha,\vv{q}}(\del_{\tau}+\omega_{\alpha,\vv{q}})a_{\alpha,\vv{q}},
\end{align}
where $a_{\alpha,\vv{q}}$ is a c-number and the $\bar{a}_{\alpha,\vv{q}}$ is its conjugate.
The action for the electron-magnon interaction, which corresponds to Eq.~(\ref{Hem}), is written as
\begin{align}
    S_{\rm int}
   & =\int d\tau 
   {
    \sum_{\boldsymbol{k}, \boldsymbol{q}}
    \sum_{m, n}
    \sum_{\alpha}}
    \ab(
    M^{\alpha,m,n}_{\vv{k},\vv{q}}a_{\alpha,\vv{q}}
    +
    M^{*\alpha,n,m}_{\vv{k}+\vv{q},-\vv{q}}\bar{a}_{\alpha,-\vv{q}}
    )
    \notag\\& \hspace{7em} \times
    \bar{\psi}_{m,\vv{k}+\vv{q}}\psi_{n,\vv{k}}.
\end{align}

The path-integral representation of the partition function is given by
\begin{align}
    Z =\int \mathcal{D}[\bar{\psi}\psi] 
        \mathcal{D}[\bar{a} a] 
        \,
        e^{-S_{\rm WPF} -S_{\rm FMI} - S_{\rm int}}.
    \label{partition_full}
\end{align}
In the following, we discuss the effective action for fermions. The partition function is divided into the boson and fermion parts as
\begin{align}
 Z &= Z_{\rm FMI} Z_{\rm eff},
 \\
 Z_{\rm eff} &= \int \mathcal D [\bar\psi  \psi] 
 e^{-S_{\rm WPF}}
 \ab<e^{-S_{\rm int}}>_{\rm FMI},
\end{align}
where we define the partition function for magnons $Z_{\rm FMI}$ as
\begin{align}
    Z_{\rm FMI} = \int\mathcal{D}[\bar{a} a]
    \,
    e^{ -S_{\rm FMI} },
\end{align}
and the expectation value of $O$ for magnons
\begin{align}
 \ab<O>_{\rm FMI} = \frac{1}{Z_{\rm FMI}}
 \int \mathcal D [\bar a  a] 
 \,
 e^{-S_{\rm FMI}} 
 O.
\end{align}
The effective action, $\ab<e^{-S_{\rm int}}>_{\rm FMI} = e^{-S_{\rm eff}}$, is given as
\begin{align}
    S_{\rm eff}
    = -\frac{1}{2}\Lb S_{\rm int}^2 \Rb_{\rm FMI}.
\end{align}
Note that the higher-order cumulants vanish for the Gaussian distribution and $\ab<S_{\rm int}>_{\rm FMI} = 0$ because of $\ab<a_{\alpha, \boldsymbol{q}}>_{\rm FMI} = 0$.
Therefore, we obtain the effective action
\begin{align}
    S_{\rm eff}
   & = \frac{1}{2}
   {
    \sum_{\boldsymbol{k}, \boldsymbol{k}', \boldsymbol{q}}
    \sum_{m,m',n,n'}}
    \int d\tau d\tau'
    V_{{\boldsymbol{k}, \boldsymbol{k}', \boldsymbol{q}}}^{{n,n';m,m'}}(\tau-\tau')
    \notag\\& \quad \times
    \bar\psi_{m, \boldsymbol{k}+\boldsymbol{q}}(\tau)
        \bar\psi_{m', \boldsymbol{k}'-\boldsymbol{q}}(\tau')
        \psi_{n', \boldsymbol{k}'}(\tau')
    \psi_{n, \boldsymbol{k}}(\tau),
\end{align}
\begin{align}
    V^{{n,n';m,m'}}_{\vv{k}\vv{k}'\vv{q}}(\tau-\tau')
    &=
	\sum_\alpha
    D_{\alpha, \boldsymbol{q}}(\tau-\tau')
    M^{\alpha,m,n}_{\vv{k},\vv{q}}
    M^{*\alpha,n',m'}_{\vv{k}'-\vv{q},\vv{q}}
\notag\\&
    +
    \sum_\alpha
    D_{\alpha, \boldsymbol{q}}(\tau'-\tau)
    M^{\alpha,m',n'}_{\vv{k}',-\vv{q}}
    M^{*\alpha,n,m}_{\vv{k}+\vv{q},-\vv{q}},
\end{align}
and we define the Fourier transformation as
\begin{align}
    \psi_{n,\vv{k}}(\tau) &= 
    \frac{1}{\sqrt{\beta}}
    \sum_{l}e^{-i\epsilon_l \tau}
    \psi_{n,\vv{k}}(i\epsilon_l),
    \\
    \bar\psi_{n,\vv{k}}(\tau)&= 
    \frac{1}{\sqrt{\beta}}
    \sum_{l}e^{i\epsilon_l \tau}
    \bar\psi_{n,\vv{k}}(i\epsilon_l),
    \\
    \mathscr{D}_{\alpha,\vv{q}}(\tau-\tau')
    &\equiv -\Lb 
    a_{\alpha,\vv{q}}(\tau) 
    \bar{a}_{\alpha,\vv{q}}(\tau')
    \Rb_{\rm FMI}
    \notag\\
    &= \frac{1}{\beta}\sum_{l}
    \frac{e^{-i\omega_{l}(\tau-\tau')}}{i\omega_l -\omega_{\alpha,\vv{q}}},
\end{align}
where $\beta$ is the inverse temperature.

The effective action for the Amperean pairing, 
has the form
\begin{align}
&
 S_{\rm eff}^{\rm Amp}
 = \frac{1}{2}
 {
 \sum_{\boldsymbol{p}, \boldsymbol{p}'}
 \sum_{m,m',n,n'}}
 \int d\tau d\tau'
 U_{\boldsymbol{p} \boldsymbol{p}' \boldsymbol{K}}^{nn',mm'}(\tau-\tau')
 \notag\\&  \times
 \bar{\psi}_{m, \boldsymbol{K}+\boldsymbol{p}}(\tau)
 \bar{\psi}_{m', \boldsymbol{K}-\boldsymbol{p}} (\tau')
 {\psi}_{n', \boldsymbol{K}-\boldsymbol{p}'}(\tau')
 {\psi}_{n, \boldsymbol{K}+\boldsymbol{p}'} (\tau),
\end{align}
with
\begin{align}
 U_{\boldsymbol{p}, \boldsymbol{p}', \boldsymbol{K}}^{n,n';m,m'}(\tau-\tau')
 = V_{\boldsymbol{K}+\boldsymbol{p}', \boldsymbol{K}-\boldsymbol{p}', \boldsymbol{p} - \boldsymbol{p}'}^{n,n';m,m'}(\tau-\tau').
\end{align}

\section{Derivation of linearized Eliashberg equation}
\label{app:Eliashberg}

Here, we derive the linearized Eliashberg equation for the Amperean pairing, 
following Refs.~\cite{goryo1998abelian, Protter_SC}.
We consider the partition function of wallpaper fermion defined as
\begin{align}
    Z_{\rm eff} 
    &= \int \mathcal{D}[\bar{\psi}\psi]
    e^{-S_{\rm WPF}-S_{\rm eff}^{\rm Amp}},
    \label{partition_WPF}
\end{align}
Now, we introduce the auxiliary field $\Phi$ and define
\begin{align}
    N &= 
    \int \mathcal{D}[\bar{\Phi}\Phi]
    e^{-\Delta S},
    \notag\\
    \Delta S
    &\equiv
    -
    \sum_{\boldsymbol{p}, \boldsymbol{p}'}
    \sum_{m,m',n,n'}
    \int d\tau d\tau'
    \frac{U^{n,n';m,m'}_{\vv{p},\vv{p}',\vv{K}}(\tau-\tau')}{2}
    \notag\\
    &\quad\times
    \lb
    \bar{\Phi}_{\vv{p}}^{mm'}(\tau-\tau')- 
    \bar{\psi}_{m, \boldsymbol{K}+\boldsymbol{p}}(\tau)
    \bar{\psi}_{m', \boldsymbol{K}-\boldsymbol{p}} (\tau')
    \rb
    \notag\\
    &\quad\times
    \lb
    \Phi_{\vv{p}'}^{nn'}(\tau-\tau') - 
    {\psi}_{n', \boldsymbol{K}-\boldsymbol{p}'}(\tau')
    {\psi}_{n, \boldsymbol{K}+\boldsymbol{p}'} (\tau)
    \rb.
\end{align}
Putting $1=\int \mathcal D[\bar\Phi\Phi] e^{-\Delta S}/N$ to the partition function Eq.~(\ref{partition_WPF}), we obtain 
\begin{align}
    Z_{\rm eff}
    &= \frac{1}{N}
    \int \mathcal{D}[\bar{\Phi}\Phi]
    \mathcal{D}[\bar{\psi}\psi]
    e^{-S_{\rm WPF}-S(\Phi)},
    \notag\\
    S(\Phi) &\equiv S_{\rm eff}^{\rm Amp} + \Delta S.
\end{align}
The action is rewritten in the Nambu form as
\begin{align}
	&
    S_{\rm WPF}+S(\Phi)
    =
    -
    \frac{1}{2}
    \sum_p
    \tilde{\psi}_p^{\dag}
    \mqty(
    {\mathscr{G}}_{0}^{-1}(p) & {\Delta}_p \\
    {\bar{\Delta}}_p & -{\mathscr{G}}_{0}^{-1}(-p)
    )
    \tilde{\psi}_p
    \notag\\
    &
    -
    \sum_{\boldsymbol{p},\boldsymbol{p}'}
    \sum_{m,m',n,n'}
    \int d\tau d\tau'
    \frac{U^{n,n';m,m'}_{\vv{p},\vv{p}',\vv{K}}(\tau-\tau')}{2}
    \notag\\& \hspace{9em} \times
    \bar{\Phi}_{\vv{p}}^{mm'}(\tau-\tau')
    \Phi_{\vv{p}'}^{nn'}(\tau-\tau'),
    \notag\\
    &
    \tilde{\psi}_{{p}} = \mqty(
    {\psi}_{1,\vv{K}+p} \\ 
    {\psi}_{2,\vv{K}+p} \\
    \bar{\psi}_{1,\vv{K}-p} \\ 
    \bar{\psi}_{2,\vv{K}-p} ),
    \
    p = (\boldsymbol{p}, i\epsilon_l),
    \notag\\
    &\tilde{\psi}_{{p}}^{\dag} = \mqty(
    \bar{\psi}_{1,\vv{K}+{p}} &
    \bar{\psi}_{2,\vv{K}+{p}} &
    {\psi}_{1,\vv{K}-{p}} &
    {\psi}_{2,\vv{K}-{p}} 
    )
\end{align}
where we define the propagator of wallpaper fermion {${\mathscr G}_{0}$} and the gap function {$\Delta$} as
\begin{align}
    ({\mathscr{G}}_{0}^{-1})^{m,m'}(p)
    &=
    (\mathscr{G}_0^{-1})^m(p)
     \delta_{mm'},
     \notag\\
     {(\mathscr{G}_0^{-1})^m(p)}
     &=
     i \epsilon_l - \xi_{m, \boldsymbol{K}+\boldsymbol{p}},
    \notag\\
    ({\Delta}_{p})_{{m,m'}}
    &=-\frac{1}{\beta}\sum_{n,n'}\sum_{p'}
    U^{n,n';m,m'}_{\vv{p},\vv{p}',\vv{K}}(i\epsilon_{l}-i\epsilon_{l'})
    \Phi_{{p}'}^{nn'},
    \notag\\
    ({\bar{\Delta}}_{p})_{{m,m'}}
    &=
    -\frac{1}{\beta}\sum_{n,n'}\sum_{p'}
    U^{m',m;n,n'}_{\vv{p}',\vv{p},\vv{K}}(i\epsilon_{l'}-i\epsilon_{l})
    \bar{\Phi}_{{p}'}^{nn'},
    \notag\\
    U^{n,n';m,m'}_{\vv{p},\vv{p}',\vv{K}}(\tau)
    &=\frac{1}{\beta}\sum_{\omega_{n}}e^{-i\omega_n \tau}U^{n,n';m,m'}_{\vv{p},\vv{p}',\vv{K}}(i\omega_n),
    \notag\\
    {\Phi}_{\vv{p}}^{mm'}(\tau) 
    &= \frac{1}{\beta}\sum_{\epsilon_{{l}}}
    e^{-i\epsilon_{{l}} \tau}
    {\Phi}_{{p}}^{mm'},
    \notag \\
    {\bar{\Phi}}_{\vv{p}}^{mm'}(\tau) 
    &= \frac{1}{\beta}\sum_{\epsilon_{{l}}}
    e^{i\epsilon_{{l}} \tau}
    {\bar{\Phi}}_{{p}}^{mm'}.
    \label{def_Delta_self}
\end{align}
Integrating out the fermions, we find the partition function for the auxiliary field as
\begin{align}
    Z_{\rm eff} =&
    \int \mathcal{D}[\bar{\Phi}\Phi]
    \exp\LB
    \frac{1}{2}\Tr\log(
    \beta
    \widetilde{G}_{p}^{-1}
    )
    \RB
    \notag\\&\hspace{-3em}\times
    \exp[
    \sum_{p,p'}
    \sum_{m,m',n,n'}
    \frac{U^{n,n';m,m'}_{\vv{p},\vv{p}',\vv{K}}
    	(i\epsilon_l-i\epsilon_{l'})}
    {2\beta}
    \bar{\Phi}_{p}^{mm'}
    \Phi_{p'}^{nn'}],
    \notag\\
    \widetilde{G}_{p}^{-1}
    =&\mqty(
    {\mathscr{G}}_{0}^{-1}(p) & {\Delta}_{p} \\
    {\bar{\Delta}}_{p} & -{\mathscr{G}}_{0}^{-1}(-p)
    ),
\end{align}

The stationary solution is determined by $\delta S/\delta \bar{\Phi}^{mm'}_{p}=0$ as
\begin{align}
    ({\Delta}_{p})_{{m,m'}}
    =
    \sum_{p'}
    \Tr \widetilde{G}_{p'}
    \mqty(
    0 & 0\\
    \displaystyle\frac{{\delta\bar{\Delta}}_{p'}}{\delta \bar{\Phi}^{mm'}_{p}} & 0
    ).
\end{align}
$\widetilde{G}_{p}$ is expanded up to the first order of ${\Delta}$ as
\begin{align}
    \widetilde{G}_{p}
    \simeq
    \mqty(
    {\mathscr{G}}_{0}(p) & {\mathscr{G}}_{0}(p){\Delta}_{p}{\mathscr{G}}_{0}(-p) 
    \\[1ex]
    {\mathscr{G}}_{0}(-p)
    {\bar{\Delta}}_{p}
    {\mathscr{G}}_{0}(p) & -{\mathscr{G}}_{0}(-p)
    ).
    \label{stationary_Phi}
\end{align}
Thus, we obtain the linearized gap equation
\begin{align}
 ({\Delta}_p)_{{m,m'}}
 =&- \frac{1}{\beta}
 \sum_{n,n'}
 \sum_{p'}
 U_{\boldsymbol{p}, \boldsymbol{p}', \boldsymbol{K}}^{n,n'; m,m'}(i\epsilon_{{l}}-i\epsilon_{{l'}})
 \notag\\&\times
 \mathscr{G}_0^{n}(p')
 \mathscr{G}_0^{n'}(-p')
 ({\Delta}_{p'})_{{n,n'}}.
\end{align}

Here, we neglect the retardation effect as
\begin{align}
	&
 U_{\boldsymbol{p}, \boldsymbol{p}', \boldsymbol{K}}^{n,n';m,m'}(i \epsilon_l - i \epsilon_{l'})
 \simeq
 U_{\boldsymbol{p}, \boldsymbol{p}', \boldsymbol{K}}^{n,n';m,m'}(0)
 \notag\\&
 = 
 -\sum_{\alpha} 
 \frac{
 	M_{\boldsymbol{K}+\boldsymbol{p}', \boldsymbol{p}-\boldsymbol{p}'}^{\alpha, m, n}
 	M_{\boldsymbol{K}-\boldsymbol{p}, \boldsymbol{p}-\boldsymbol{p}'}^{*\alpha, n', m'}
 	}
 	{\omega_{\alpha, \boldsymbol{p}-\boldsymbol{p}'}}
 \notag\\&\quad
 -
 \sum_{\alpha}
 \frac{
 	M_{\boldsymbol{K}-\boldsymbol{p}', \boldsymbol{p}'-\boldsymbol{p}}^{\alpha, m', n'}
 	M_{\boldsymbol{K}+\boldsymbol{p}, \boldsymbol{p}'-\boldsymbol{p}}^{*\alpha, n, m}}
 	{\omega_{\alpha, \boldsymbol{p}-\boldsymbol{p}'}}.
    \label{app_E_eff_int_0}
\end{align}
The gap function becomes independent of the frequency for the frequency-independent interaction, $(\Delta_p)_{m,m'} \simeq \Delta_{m,m'}(\boldsymbol{p})$.
Thus, the remaining Matsubara sum is as follows.
\begin{align}
	&
\frac{1}{\beta}
 \sum_{\epsilon_{l'}}
 \mathscr{G}_0^{n}(p)
 \mathscr{G}_0^{n'}(-p')
 \notag\\&
 = 
 \frac{1-f(\xi_{n, \boldsymbol{K}+\boldsymbol{p}'}) -f(\xi_{n', \boldsymbol{K}-\boldsymbol{p}'})}{\xi_{n, \boldsymbol{K}+\boldsymbol{p}'} + \xi_{n', \boldsymbol{K}-\boldsymbol{p}'}}
 =: \chi_{\boldsymbol{p}', \boldsymbol{K}}^{n, n'}.
\end{align}
Therefore, the gap equation reduces to
\begin{align}
 \Delta_{m,m'}(\boldsymbol{p})
 = - \sum_{n,n'}
 \sum_{\boldsymbol{p}'}
 U_{\boldsymbol{p}, \boldsymbol{p}', \boldsymbol{K}}^{n,n';m,m'}(0)
 \chi_{\boldsymbol{p}', \boldsymbol{K}}^{n, n'}
 \Delta_{n,n'}(\boldsymbol{p}').
\end{align}
Note that the gap equation for the BCS pairing with no net momentum is obtained by susbstituting $\boldsymbol{K}=\boldsymbol{0}$ as
\begin{align}
 \Delta_{m,m'}(\boldsymbol k) 
 = - \sum_{n,n'}
 \sum_{\boldsymbol{k}'}
 V_{\boldsymbol{k}, \boldsymbol{k'}}^{n,n';m,m'}
 \chi_{\boldsymbol{k'}}^{n,n'}
 \Delta_{n,n'}(\boldsymbol{k'}),
\end{align}
and
\begin{align}
 V_{\boldsymbol{k}, \boldsymbol{k'}}^{n,n'; m,m'}
 &= U_{\boldsymbol{k},\boldsymbol{k'}, \boldsymbol{0}}^{n,n';m,m'}(0)
 \notag\\&
 = -\sum_{\alpha} 
 \frac{
 	M_{\boldsymbol{k}', \boldsymbol{k}-\boldsymbol{k}'}^{\alpha, m, n}
 	M_{-\boldsymbol{k}, \boldsymbol{k}-\boldsymbol{k}'}^{*\alpha, n', m'}
 	}
 	{\omega_{\alpha, \boldsymbol{k}-\boldsymbol{k}'}}
 \notag\\&\quad
 -
 \sum_{\alpha}
 \frac{
 	M_{-\boldsymbol{k}', \boldsymbol{k}'-\boldsymbol{k}}^{\alpha, m', n'}
 	M_{\boldsymbol{k}, \boldsymbol{k}'-\boldsymbol{k}}^{*\alpha, n, m}}
 	{\omega_{\alpha, \boldsymbol{k}-\boldsymbol{k}'}}.
\end{align}

\section{Effective interaction in the isotropic limit}\label{app_eff_int_Amp}
\label{App_isotoropic}

We show a detailed derivation of the effective interaction Eq.~(\ref{eq:isotropic-limit}) near the Fermi surface, $\boldsymbol{p}$, $\boldsymbol{p} \sim \boldsymbol{0}$, in the isotropic limit for the Amperean pairing. 
The electron-magnon coupling constant Eqs.~(\ref{eq:M1})--(\ref{eq:M4}) is summarized as 
	\begin{align}
	M_{\boldsymbol{K}, \boldsymbol{0}}^{\beta, m, n}
	&=
	\frac{\tilde V}{2 \sqrt{2}}
	\ab(e^{-i\phi} -\beta mn e^{i\phi})
	\notag\\&\quad\times
	\ab|v_m^{+*}(\boldsymbol{K})
	u_n^{+}(\boldsymbol{K})|
	e^{-i \varphi_{m, \boldsymbol{K}}}.
\end{align}
Then we find
\begin{align}
 &
	M_{\vv{K}-\vv{p}',\vv{p}'-\vv{p}}^{\beta,n',m'} M_{\vv{K}+\vv{p},\vv{p}'-\vv{p}}^{*\beta,m,n}
	\simeq
	M_{\vv{K},\vv{0}}^{\beta,n',m'}
	M_{\vv{K},\vv{0}}^{*\beta,m,n}
	\notag\\&
	= \frac{\tilde V^2}{2}
	B_{\beta, n'm', mn}
	X_{\boldsymbol{K}}^{n', m', m, n}
	e^{-i (
		\varphi_{n', \boldsymbol{K}}
		-
		\varphi_{m, \boldsymbol{K}}
		)
	},
\end{align}
\begin{align}
 B_{\beta, \beta, \beta} 
 &= \sin^2\phi,
 \\
 B_{\beta, \mp\beta, \pm\beta} 
 &= \pm i \sin\phi\cos\phi,
 \\
 B_{\beta, -\beta, -\beta} 
 &= \cos^2\phi,
	\label{phase_band}
\end{align}
and
\begin{align}
	X_{\boldsymbol{K}}^{n', m', m, n}
	= \ab|v_{n'}^+(\boldsymbol{K})
	u_{m'}^+(\boldsymbol{K})
	v_{m}^+(\boldsymbol{K})
	v_{n}^+(\boldsymbol{K})
	|.
\end{align}
Here, we assign $m=+1$ for the $m=1$ band and $m=-1$ for the $m=2$ band. 

Now we consider the isotropic limit $v_{s1}=0$, in which the following holds
\begin{align}
	&\xi_{+,\vv{k}} = \xi_{-,\vv{k}},
	\quad
	\varphi_{+} = \varphi_{-},
	\quad
	u_{+}^{+} = u^{+}_{-}, 
	\quad
	v_{+}^{+} = v^{+}_{-},
	\notag\\
	&X^{n',m',m,n}_{\vv{K}} \equiv X_{\vv{K}}.
\end{align}
From the above relations and Eq.~(\ref{eq:denom}), the interaction matrix $U^{m,m';n,n'}_{\vv{p},\vv{p}',\vv{K}}$ is evaluated as
\begin{align}
	&U^{m,m';n,n'}_{\vv{p},\vv{p}',\vv{K}}
	\notag\\
	&\simeq
	-\frac{
		M^{\alpha,n,m}_{\vv{K}, \boldsymbol{0}}
		M^{*\alpha,m',n'}_{\vv{K},\vv{0}}
		+
		M^{\alpha,n',m'}_{\vv{K}, \vv{0}}
		M^{*\alpha,m,n}_{\vv{K}, \vv{0}}
	}{\omega_{+,\vv{p}-\vv{p}'}}
	\notag\\
	&\simeq
	-\frac{
		\tilde{V}^2
		\lb
		B_{+,nm,m'n'} + B_{+,n'm',mn}
		\rb
		X_{\vv{K}}
	}{2\omega_{+,\vv{p}-\vv{p}'}}
	\notag\\
	&=
	-\frac{\tilde{V}^2 X_{\vv{K}}}{\omega_{+,\vv{p}-\vv{p}'}}
	\times
	\begin{cases}
		\cos^2 \phi & mn = m'n' = 1,
		\\
		\sin^2 \phi & mn = m'n' = -1,
		\\
		0 & \text{otherwise}.
	\end{cases}
\end{align}
Next, we evaluate the antisymmetrized interaction matrix as
\begin{widetext}
\begin{equation}
U^{m,m';n,n'}_{O,\vv{p},\vv{p}',\vv{K}}
\simeq
	-\frac{\tilde{V}^2 X_{\vv{K}}}{2}\times
	\begin{cases}
		\displaystyle\frac{\cos^2 \phi}{\omega_{+,\vv{p}-\vv{p}'}} 
		-\frac{\cos^2 \phi}{\omega_{+,\vv{p}+\vv{p}'}}
		,
		& m=m'=n=n',
		\\[1em]
		\displaystyle\frac{\sin^2 \phi}{\omega_{+,\vv{p}-\vv{p}'}} 
		-\frac{\sin^2 \phi}{\omega_{+,\vv{p}+\vv{p}'}}
		,
		& m=m' = -n= -n',
		\\[1em]
		\displaystyle\frac{\cos^2 \phi}{\omega_{+,\vv{p}-\vv{p}'}} 
		-\frac{\sin^2 \phi}{\omega_{+,\vv{p}+\vv{p}'}},
		& m=-m'=n=-n',
		\\[1em]
		\displaystyle\frac{\sin^2 \phi}{\omega_{+,\vv{p}-\vv{p}'}} 
		-\frac{\cos^2 \phi}{\omega_{+,\vv{p}+\vv{p}'}},
		&m=-m'=-n=n',
		\\[1em]
		0, & \text{otherwise}.
	\end{cases}
\end{equation}
\end{widetext}
Consequently, we arrive at Eq.~(\ref{eq:isotropic-limit})
\begin{widetext}
\begin{equation}
	U^{m,m';n,n'}_{O,\vv{p},\vv{p}',\vv{K}}
		\simeq
		-\frac{\tilde{V}^2 X_{\vv{K}}}{2}\times
		\begin{cases}
			\displaystyle\frac{J_{1}\cos^2 \phi}{2sK^2}\vv{p}\cdot\vv{p}'
			,
			& m=m'=n=n',
			\\[1em]
			\displaystyle\frac{J_{1}\sin^2 \phi}{2sK^2}\vv{p}\cdot\vv{p}'
			,
			& m=m'=-n=-n',
			\\[1em]
			\displaystyle\frac{\cos 2\phi}{2sK} 
			+ \frac{J_{1}\vv{p}\cdot\vv{p}'}{4sK^2},
			& m=-m'=n=-n',
			\\[1em]
			-\displaystyle\frac{\cos 2\phi}{2sK} 
			+ \frac{J_{1}\vv{p}\cdot\vv{p}'}{4sK^2},
			& m=-m'=-n=n',
			\\[1em]
			0, & \text{otherwise}.
		\end{cases}
	\end{equation}
\end{widetext}

\bibliography{MMSC_FMWPF}
\clearpage
\end{document}